# Superconducting properties of tin-based ENZ and hyperbolic metamaterials

Vera N. Smolyaninova[1)], William Korzi[1)], William Zimmerman[1)], Sabrina Searfoss[1)], Christopher Jensen[1)], Grace Yong[1)], David Schaefer[1)], Joseph C. Prestigiacomo[2)], M. S. Osofsky[2)], Heungsoo Kim[2)], Zhen Xing[3)], M. M. Qazilbash[3)], and Igor I. Smolyaninov[4,5)]

[1]*Department of Physics Astronomy and Geosciences, Towson University,*

*8000 York Rd., Towson, MD 21252, USA*

[2] *Naval Research Laboratory, Washington, DC 20375, USA*

[3]*Department of Physics, College of William and Mary, Williamsburg, Virginia 23187-8795, USA*

[4] *Department of Electrical and Computer Engineering, University of Maryland, College Park, MD 20742, USA*

[5] *Saltenna LLC, 1751 Pinnacle Drive #600 McLean, VA 22102, USA*

**Recent experiments have demonstrated that the superconducting critical temperature may be improved in various metamaterial superconductor geometries. Here, we present the results of a study of tin-based metamaterial superconductors in the epsilon-near-zero (ENZ) and hyperbolic metamaterial configurations. It was observed that $T_c$ enhancement is significantly reduced when the metamaterial structural dimensions exceed 240nm, the superconducting coherence length in pure tin.**

## 1. Introduction

Designing new materials with enhanced superconducting properties remains one of the most important goals of condensed matter physics. A strategy that deliberately engineers the dielectric properties of nanostructured composite superconductor-dielectric metamaterials has been demonstrated to result in an enhanced electron pairing interaction that increases the value of the critical temperature, $T_c$, of so-called "metamaterial superconductors" [1-5]. For example, $Al_2O_3$-coated aluminum nanoparticles can be used to form epsilon near zero (ENZ) core-shell metamaterial superconductors with a $T_c$ that is three times that of pure aluminum [4]. Alternatively, a multilayer $Al/Al_2O_3$ hyperbolic metamaterial structure may be fabricated [5], which exhibits similar $T_c$ enhancement, while having much better transport properties. Hyperbolic metamaterials are extremely anisotropic uniaxial materials, which behave like a metal ($\varepsilon_{xx} = \varepsilon_{yy} < 0$) in one direction and like a dielectric ($\varepsilon_{zz} > 0$) in the orthogonal direction [6]. The hyperbolic metamaterial geometry shown in Fig. 1 is a structure based on periodic metallic layers separated by dielectric layers. As was originally noted in [7], typical high $T_c$ superconductors (such as BSCCO) exhibit hyperbolic behavior in the far infrared and THz frequency ranges, which may be partially responsible for their very large $T_c$ values.

As has been noted above, most of the initial metamaterial superconductor geometries were successfully demonstrated using aluminum as a superconducting component of the composite metamaterial. Aluminum is ideal for these proof of principle experiments because its critical temperature is quite low ($T_{c\ Al}$ = 1.2 K [8]), leading to a very large superconducting coherence length $\xi$ = 1600 nm [8]. Such a large value of $\xi$ facilitates the metamaterial fabrication requirements since the validity of macroscopic electrodynamic description of the metamaterial superconductor requires that its spatial structural parameters must be much smaller than $\xi$. In this paper we will expand this approach to another elemental superconductor, tin, which also has rather large superconducting coherence length $\xi$ = 240 nm [8]. We have realized both ENZ and hyperbolic metamaterial geometries based on tin, and observed the effect of $T_c$ increase in the tin-based metamaterial superconductors. In addition, we studied the effect of metamaterial component dimensions on $T_c$ in the tin-based ENZ metamaterials demonstrating that $T_c$ enhancement is significantly reduced when the metamaterial structural parameter exceeds 240 nm, the coherence length of pure tin. This observation supports the validity of the macroscopic electrodynamic (metamaterial) approach to designing new materials with enhanced superconducting properties.

## 2. Macroscopic electrodynamic description of superconductivity

Electromagnetic properties play a very important role in the electron pairing in superconductors [9]. According to the BCS theory, the Cooper pairing mechanism may be described as an attractive interaction of electrons which results from the polarization of the ionic lattice created by motion of these electrons through the lattice. Based on this physical picture, Kirzhnits *et al.* formulated their description of superconductivity in terms of the dielectric response function of the superconductor [9]. Within the scope of this “macroscopic electrodynamics” approach, the electron-electron interaction in a superconductor can be expressed in the form of an effective Coulomb potential

$$V(\vec{q},\omega)=\frac{4\pi e^2}{q^2\varepsilon_{eff}(\vec{q},\omega)}=V_C\frac{1}{\varepsilon_{eff}(\vec{q},\omega)} \quad (1)$$

where $V_C$ is the Fourier-transformed Coulomb potential in vacuum, and $\varepsilon_{eff}(q,\omega)$ is the dielectric response function of the superconductor, which is treated as an effective medium. Following this formalism, it appears natural to use the recently developed electromagnetic metamaterial toolbox [10] to engineer the $\varepsilon_{eff}(q,\omega)$ of an artificial "metamaterial superconductor", so that the electron pairing interaction described by Eq.(1) will be maximized. Indeed, it was predicted in [1, 2] that a considerable enhancement of the electron pairing interaction may be expected in the ENZ [11] and hyperbolic metamaterial scenarios [6]. In both cases $\varepsilon_{eff}(q,\omega)$ may become small and negative in substantial portions of the four-momentum $(q,\omega)$ space near the electron Fermi surface, leading to considerable enhancement of the electron pairing interaction. For example, in the case of a hyperbolic metamaterial the effective Coulomb potential from Eq. (1) assumes the form

$$V(\vec{q},\omega)=\frac{4\pi e^2}{q_z^2\varepsilon_2(\vec{q},\omega)+\left(q_x^2+q_y^2\right)\varepsilon_1(\vec{q},\omega)} \quad (2)$$

where $\varepsilon_{xx} = \varepsilon_{yy} = \varepsilon_1$ and $\varepsilon_{zz} = \varepsilon_2$ have opposite signs. As a result, the effective Coulomb interaction of two electrons may become attractive and very strong along spatial directions where

$$q_z^2\varepsilon_2(\vec{q},\omega)+\left(q_x^2+q_y^2\right)\varepsilon_1(\vec{q},\omega)\approx 0 \quad (3)$$

As summarized in [12], the critical temperature of a superconductor in the weak coupling limit is typically calculated as

$$T_c = \theta \ \exp\left(-\frac{1}{\lambda_{eff}}\right), \tag{4}$$

where $\theta$ is the characteristic temperature for a bosonic mode mediating electron pairing (such as the Debye temperature $\theta_D$ in the standard BCS theory), and $\lambda_{eff}$ is the dimensionless coupling constant defined by $V(q,\omega) = V_C(q)\ \varepsilon^{-1}(q,\omega)$ and the density of states $\nu$ (see for example [13]):

$$\lambda_{eff} = -\frac{2}{\pi}\nu\int_0^{\infty}\frac{d\omega}{\omega}\left\langle V_C(q)\,\mathrm{Im}\,\varepsilon^{-1}(\vec{q},\omega)\right\rangle, \tag{5}$$

where $V_C$ is the unscreened Coulomb repulsion, and the angle brackets denote average over the Fermi surface. Additional zeroes of the effective dielectric response function $\varepsilon_{eff}(q,\omega)$ of the metal-dielectric metamaterial described by Eq. (3) lead to increase of $\lambda_{eff}$ compared to the bulk metal, resulting in higher $T_c$ of the metamaterial. Very recently it was demonstrated that a multilayer $Al/Al_2O_3$ hyperbolic metamaterial structure indeed exhibits considerable $T_c$ enhancement compared to the critical temperature in bulk aluminum [5]. Here, we report that similar ENZ and hyperbolic metamaterial configurations increase $T_c$ in tin-based metamaterial superconductors.

**3. Tin-based ENZ superconductors: metamaterial structural parameter size effect**

According to the Maxwell-Garnett approximation [14], mixing of nanoparticles of a superconducting "matrix" with dielectric "inclusions" (described by the dielectric constants $\varepsilon_m$ and $\varepsilon_d$, respectively) results in the effective medium with a dielectric constant $\varepsilon_{eff}$, which may be obtained as

$$\left(\frac{\varepsilon_{eff}-\varepsilon_m}{\varepsilon_{eff}+2\varepsilon_m}\right)=(1-n)\left(\frac{\varepsilon_d-\varepsilon_m}{\varepsilon_d+2\varepsilon_m}\right), \qquad (6)$$

where $n$ is the volume fraction of metal. An explicit expression for the inverse dielectric response function $\varepsilon_{\mathrm{eff}}^{-1}$ of this effective medium may be written as

$$\varepsilon_{eff}^{-1}=\frac{n}{(3-2n)}\frac{1}{\varepsilon_m}+\frac{9(1-n)}{2n(3-2n)}\frac{1}{(\varepsilon_m+(3-2n)\varepsilon_d/2n)} \quad , \qquad (7)$$

which makes it apparent that $\varepsilon_{\mathrm{eff}}^{-1}$ of the mixture acquires an additional pole at

$$\varepsilon_m \approx -\frac{3-2n}{2n}\varepsilon_d \quad , \qquad (8)$$

(in addition to the bulk metal plasmon pole at $\varepsilon_m \approx 0$). This additional pole, which has been recently identified as a plasmon-phonon pole of the metal-dielectric mixture [15], contributes to the increased coupling constant $\lambda_{\mathrm{eff}}$ of the metamaterial, which leads to increased $T_c$.

The ENZ behavior described above, and the corresponding increase of the superconducting critical temperature has been recently observed in Sn/$BaTiO_3$ nanoparticle mixtures [3]. The metamaterial samples in these experiments were prepared using commercially available tin and barium titanate nanoparticles obtained from the US Research Nanomaterials, Inc. The average diameter of the $BaTiO_3$ nanoparticles was 50 nm, while tin nanoparticle size was specified as 60-80 nm. Since both nanoparticle sizes fall substantially below the superconducting coherence length in pure tin $\xi_{\mathrm{Sn}}$ ~230 nm [8], the metamaterial approach seems to be well justified in this case. The Sn/$BaTiO_3$ superconducting metamaterials were fabricated by combining the given amounts of Sn and $BaTiO_3$ nanoparticle powders by volume into a single test tube filled with de-ionized water. The resulting suspensions were sonicated and magnetically stirred for 30 minutes, followed by water evaporation. The remaining mixtures of Sn and $BaTiO_3$ nanoparticles were compressed into test pellets using a hydraulic press. The ENZ conditions in these initial experiments were achieved in the 30-50% range of the volume fraction of $BaTiO_3$. The $T_c$ increased from the pure Sn value of 3.68 K with increasing $BaTiO_3$ concentration to a maximum $\Delta T_c$ ~ 0.15 K (or 4%) compared to the pure tin sample for the 40% sample followed by $T_c$ decreasing at higher volume fractions [3].

The commercial availability of a broad variety of tin and $BaTiO_3$ nanoparticle sizes (in the 50 nm to several micrometers range) allowed us to further extend this study to larger nanoparticle sizes, and verify that the observed $T_c$ increase indeed gradually diminishes with an increase in particle sizes beyond the superconducting coherence length in pure tin $\xi_{Sn}$ ~230 nm. These experiments provide a very strong support for the metamaterial superconductor concept.

A similar metamaterial fabrication technique described above has been applied to samples containing larger nanoparticle sizes. Typical Scanning Electron Microscopy (SEM) images of the resulting Sn/$BaTiO_3$ composite metamaterials are shown in Fig. 2(a). The original compressed nanoparticles are clearly visible in both images. The resulting average metamaterial composition has been verified after fabrication using an SEM with energy dispersive X-ray spectroscopy (EDS). An example of such compositional analysis spectra is shown in Fig.2(b). The measured individual element abundancies were consistent with this nominal composition. Such EDX spectra were used to establish the homogeneous character of the fabricated composite metamaterials.

The ENZ character of the fabricated Sn/$BaTiO_3$ metamaterial superconductor samples has been verified by infrared reflectance measurements of the samples in an extended frequency and temperature range. The real and imaginary parts of the dielectric function were obtained by Kramers-Kronig analysis of the infrared reflectance data constrained by ellipsometry data at higher frequencies [16]. As plotted in Fig. 3, the real part of the frequency dependent dielectric constant of the metamaterial samples at 5 K and at room temperature indeed appeared to be much smaller in magnitude compared to the real part of the dielectric constant of tin.

The superconducting critical temperature $T_c$ of various Sn/$BaTiO_3$ metamaterial samples was measured as a function of nanoparticle sizes and volume fractions of $BaTiO_3$ via the onset of diamagnetism using a MPMS SQUID magnetometer. The zero field cooled (ZFC) magnetization per unit mass for several samples with varying concentrations of $BaTiO_3$ for two different sets of nanoparticle sizes is plotted in Fig. 4. The effect of the superconducting critical temperature $T_c$ increase appears to be strongly suppressed in metamaterial superconductors made with particle sizes considerably larger than the superconducting coherence length in pure tin $\xi_{Sn}$ ~230 nm.

The temperatures of the onset of the superconducting transition for samples made with various nanoparticle sizes as a function of $BaTiO_3$ volume fraction are plotted in Fig. 5(a). The pure tin samples were prepared from pressed tin nanoparticles of the same diameter using the same method of preparation. Note that the value of $T_c$ for these samples agreed with the expected value of pure Sn. The increase of $T_c$ for the samples with 40% volume fraction of $BaTiO_3$ measured as a function of nanoparticle size is plotted in Fig. 5(b). Once again, $T_c$ increase appears to be strongly suppressed in metamaterial superconductors made with particle sizes considerably larger than the superconducting coherence length in pure tin. These results validate the metamaterial approach to dielectric response engineering in metamaterial superconductors.

## 4. Tin-based hyperbolic metamaterial superconductors

As has been mentioned above, the hyperbolic metamaterial strategy also appears to be a very effective tool of dielectric response engineering in metamaterial superconductors. We were able to validate this approach with tin-based hyperbolic metamaterials. In the hyperbolic metamaterial geometry shown schematically in Fig.1 the role of metal layers was played by thin tin films, while either $CaF_2$ or $MgF_2$ was used as the dielectric layer material. The individual layers of Sn and $CaF_2$ were fabricated sequentially using thermal evaporation. This process was repeated to create multilayer Sn-$CaF_2$ hyperbolic metamaterial samples on glass substrates. For the metamaterial approach to be successful, the thicknesses of superconductor and dielectric layers should be much smaller than the superconducting coherence length in pure tin $\xi_{Sn}$ ~230 nm, and the dielectric layer thickness must be thin enough to allow electron tunneling between the tin layers. The increase of $T_c$ in hyperbolic metamaterials also depends on the volume fraction of metal in metal/dielectric multilayers [5]. Therefore, we tried multiple layer deposition thicknesses to find the optimal tin/$CaF_2$ ratio in the metamaterial. The actual thicknesses of the tin and $CaF_2$ films in the nanofabricated samples were then determined by Atomic Force Microscopy (AFM).

To demonstrate that our multilayer samples exhibit hyperbolic behaviour, we studied their optical properties using polarization reflectometry (Fig. 6). The theoretical values of the $\varepsilon_1$ and $\varepsilon_2$ components of the Sn/$CaF_2$ multilayer films may be obtained using the Maxwell-Garnett approximation as follows:

$$\varepsilon_1 = n\varepsilon_m + (1-n)\varepsilon_d \tag{9}$$

$$\varepsilon_2 = \frac{\varepsilon_m \varepsilon_d}{(1-n)\varepsilon_m + n\varepsilon_d} \tag{10}$$

where $n$ is the volume fraction of metal, and $\varepsilon_m$ and $\varepsilon_d$ are the dielectric permittivities of the metal and dielectric, respectively [17]. The signs of the real parts of $\varepsilon_1$ and $\varepsilon_2$ (and therefore the hyperbolic character of our samples) were determined by polarization reflectometry. The metamaterial parameters were extracted from the polarization reflectometry data as described in detail in [18]. Reflectivity for s-polarization is given in terms of the incident angle $\theta$ by

$$R_s = \left| \frac{\sin(\theta - \theta_{ts})}{\sin(\theta + \theta_{ts})} \right|^2 , \tag{11}$$

where

$$\theta_{ts} = \arcsin\left( \frac{\sin\theta}{\sqrt{\varepsilon_1}} \right) . \tag{12}$$

Reflectivity for p-polarization is given as

$$R_p = \left| \frac{\varepsilon_1 \tan\theta_{tp} - \tan\theta}{\varepsilon_1 \tan\theta_{tp} + \tan\theta} \right|^2 , \tag{13}$$

where

$$\theta_{tp} = \arctan\sqrt{\frac{\varepsilon_2 \sin^2\theta}{\varepsilon_1\varepsilon_2 - \varepsilon_1 \sin^2\theta}} . \tag{14}$$

We measured p- and s- polarized absolute reflectance of the metamaterial samples at three different laser light frequencies (488 nm, 633 nm and 1550 nm.) as indicated in

Fig. 6. The absolute reflectance values were obtained by normalization to the measured reflectance of a 150 nm gold film. The dashed lines in Fig. 6 are fits using Eqs. (11-14). $\varepsilon_1$ and $\varepsilon_2$ values obtained from the fits confirm hyperbolic character of the 8 layer $Sn/CaF_2$ metamaterial which consists of 5.3 nm Sn layers separated by 1 nm layers of $CaF_2$. It is clear that the real part of the out-of-plane dielectric function is positive, while the real part of the in-plane dielectric function is negative, which confirms the dielectric nature along z-axis and metallic nature in the xy-plane, i.e. a hyperbolic metamaterial. We have also verified that the results of these measurements demonstrate good agreement with theoretical predictions obtained from Maxwell-Garnett approximation, Eqs. (9, 10), and the reported optical properties of tin [19] and $CaF_2$ [20]. Comparison of the calculated and measured values of the real and imaginary parts of $\varepsilon_1$ and $\varepsilon_2$ is presented in Fig.7.

The $T_c$ of various samples (Figs. 8, 9) were determined via four-point resistivity measurements as a function of temperature using a Physical Property Measurement System (PPMS) by Quantum Design. Fig. 8 shows measured resistivity as a function of temperature for several multilayer $Sn/CaF_2$ metamaterials having different thicknesses of Sn layers, while the 1 nm thickness of $CaF_2$ layers was kept approximately constant. The $Sn/CaF_2$ hyperbolic metamaterial samples were found to have a higher critical temperature than pure Sn, which strongly indicates the role of hyperbolic geometry in $T_c$ enhancement.

We have also studied changes in $T_c$ as a function of tin volume fraction in a set of several $Sn/CaF_2$ and $Sn/MgF_2$ metamaterial samples, as shown in Fig. 9. The quantitative behaviour of $T_c$ as a function of $n$ may be predicted based on the hyperbolic enhancement of the electron-electron interaction (Eq. (2)) and the density of electronic states, $\nu$, on the Fermi surface which experience this hyperbolic enhancement [5]. Using Eqs. (9, 10), the effective Coulomb potential from Eq. (2) may be re-written as

$$V(\vec{q},\omega)=\frac{4\pi e^2}{q^2\left(\frac{q_z^2}{q^2}\frac{\varepsilon_d\varepsilon_m}{((1-n)\varepsilon_m+n\varepsilon_d)}+\frac{q_x^2+q_y^2}{q^2}(n\varepsilon_m+(1-n)\varepsilon_d)\right)}\ . \qquad (15)$$

Based on Eq. (15), the differential of the product $\nu V$ may be written as

$$d(\nu V)=\frac{4\pi e^2 n\sin\theta d\theta}{q^2\left(\frac{\varepsilon_d\varepsilon_m}{((1-n)\varepsilon_m+n\varepsilon_d)}\cos^2\theta+(n\varepsilon_m+(1-n)\varepsilon_d)\sin^2\theta\right)}=$$

$$=-\frac{4\pi e^2 n dx}{q^2\left((n\varepsilon_m+(1-n)\varepsilon_d)-\frac{n(1-n)(\varepsilon_m-\varepsilon_d)^2}{(n\varepsilon_d+(1-n)\varepsilon_m)}x^2\right)}\ , \qquad (16)$$

where $x = \cos\theta$, and $\theta$ varies from 0 to π. The latter expression has two poles at

$$\varepsilon_m=\left(\left(1-\frac{1}{2n(1-n)(1-x^2)}\right)\pm\sqrt{\left(1-\frac{1}{2n(1-n)(1-x^2)}\right)^2-1}\right)\varepsilon_d \qquad (17)$$

As the volume fraction, $n$, of metal is varied, one of these poles remains close to $\varepsilon_m = 0$, while the other is observed at larger negative values of $\varepsilon_m$. This behaviour is similar to the appearance of an additional pole in ENZ metamaterials (see Eqs. (7, 8)). Using Eqs.(4, 5), and following the numerical integration procedure described in detail in [5], a theoretical prediction of $T_c$ as a function of $n$ in the Sn/$CaF_2$ hyperbolic metamaterials may be obtained. This behaviour is plotted in Fig. 9 as a red curve. This figure demonstrates that the experimentally measured behaviour of $T_c$ as a function of $n$ (which is defined by the Sn layer thickness) correlates well with the theoretical prediction. As discussed in detail in [21], smaller relative increases of $T_c$ in tin-based metamaterial superconductors compared to similar aluminium-based metamaterial geometries may be explained by considerably larger imaginary part $\varepsilon_m$'' of the tin dielectric constant.

**5. Conclusions**

Our experimental results confirm recent theoretical predictions and earlier experimental observations that the superconducting critical temperature may be improved in various metamaterial superconductor geometries. Our detailed study of tin-based metamaterial superconductors in the epsilon-near-zero (ENZ) and hyperbolic metamaterial configurations demonstrated that the metamaterial superconductor approach remains valid even as the superconductor coherence length is strongly reduced compared to the very large value of $\xi = 1600$ nm in aluminum. As expected, it was observed that the $T_c$ enhancement effect is significantly reduced when the dimensions of the metamaterial components exceed $\xi = 240$ nm superconducting coherence length in pure tin.

Our experimental observations open up numerous other possibilities for considerable $T_c$ enhancement in other practically important simple superconductors. However, due to their much smaller coherence length metamaterial structuring of these superconductors must be performed on a much more refined scale.

This work was supported in part by the DARPA Award No: W911NF-17-1-0348 "Metamaterial Superconductors", and ONR Awards N00014-18-1-2681 and N00014-18-1-2653. M.M.Q. acknowledges support from the National Science

Foundation, Division of Materials Research (Award No. NSF DMR-1255156) for the research conducted at William and Mary.

**Figure Captions**

**Fig. 1.** Schematic geometry of a "layered" hyperbolic metamaterial. In the present work tin was used as metallic layers, while either $CaF_2$ or $MgF_2$ were used as the dielectric layers.

**Fig. 2**. (a) SEM images of the composite Sn/$BaTiO_3$ metamaterials. Individual compressed nanoparticles are clearly visible in the images. The sample compositions and Sn and $BaTiO_3$ nanoparticle sizes are indicated above the images. (b) Verification of the average metamaterial composition using an SEM with energy dispersive X-ray spectroscopy (EDS).

**Fig. 3**. Confirmation of the ENZ behavior of the Sn/$BaTiO_3$ metamaterial superconductor samples: The real part, $\varepsilon'$, of the dielectric constant of the metamaterial samples at 5 K and at room temperature appears to be much smaller in magnitude compared to the real part of the dielectric constant of tin.

**Fig. 4**. Temperature dependence of zero field cooled magnetization per unit mass measured in a magnetic field of 10 G for several tin-$BaTiO_3$ metamaterial samples with varying volume fraction of $BaTiO_3$ (indicated in the plot) for samples made with (a) 70 nm tin and 50 nm $BaTiO_3$ nanoparticles (b) 1000 nm tin and 3000 nm $BaTiO_3$ nanoparticles.

**Fig. 5**. (a) The temperatures of the onset of the superconducting transition plotted as a function of volume fraction of $BaTiO_3$ for samples made with various nanoparticle sizes. The lines are a guide to the eye. (b) The increase of $T_c$ for the samples with 40% volume fraction of $BaTiO_3$ measured as a function of nanoparticle size. The superconducting critical temperature $T_c$ increase appears to be strongly suppressed in metamaterial superconductors made with particle sizes considerably larger than the superconducting coherence length in pure tin $\xi_{Sn}$ ~230 nm.

**Fig**. **6**. Polarization reflectometry of a hyperbolic metamaterial consisting of 8 layers of (Sn 5.3 nm)/($CaF_2$ 1 nm). Data points are the measured p- and s-polarized reflectivities of the metamaterial sample at 488 nm, 633 nm and 1550 nm. The lines are fits using Eqs. (11-14). The fitting parameters, the real and imaginary parts of $\varepsilon_1$ and $\varepsilon_2$, are specified in the plot. The negative sign of the real part of $\varepsilon_1$ and the positive sign of the real part of $\varepsilon_2$ obtained from the fits confirm hyperbolic character of the metamaterial.

**Fig. 7**. The calculated plots of the real and imaginary parts of (a) $\varepsilon_1$ and (b) $\varepsilon_2$ of the multilayer Sn/$CaF_2$ metamaterial. The calculations were performed using Eqs. (9, 10) based on the Maxwell-Garnett approximation. The metamaterial appears to be hyperbolic in the visible and near IR ranges. The experimentally measured data points from Fig. 6 are also shown for the comparison.

**Fig**. **8**. Measured resistivity as a function of temperature for several multilayer Sn/$CaF_2$ metamaterials having different thicknesses of Sn layers, while the 1 nm thickness of $CaF_2$ layers was kept approximately constant. The thicknesses of Sn/$CaF_2$ layer are indicated. The Sn/$CaF_2$ hyperbolic metamaterial samples were found to have a higher critical temperature than pure Sn, which strongly indicates the role of the hyperbolic geometry in $T_c$ enhancement.

**Fig. 9**. Effect of the tin volume fraction $n$ on $T_c$ of the Sn/$CaF_2$ and Sn/$MgF_2$ hyperbolic metamaterial samples. Experimentally measured behavior of $T_c$ as a function of $n$ (which is defined by the tin layer thickness) correlates well with the theoretical fit (red curve) based on the hyperbolic mechanism of $T_c$ enhancement. Experimental data

points shown in black correspond to Sn/$CaF_2$ samples, while blue ones correspond to Sn/$MgF_2$ samples.

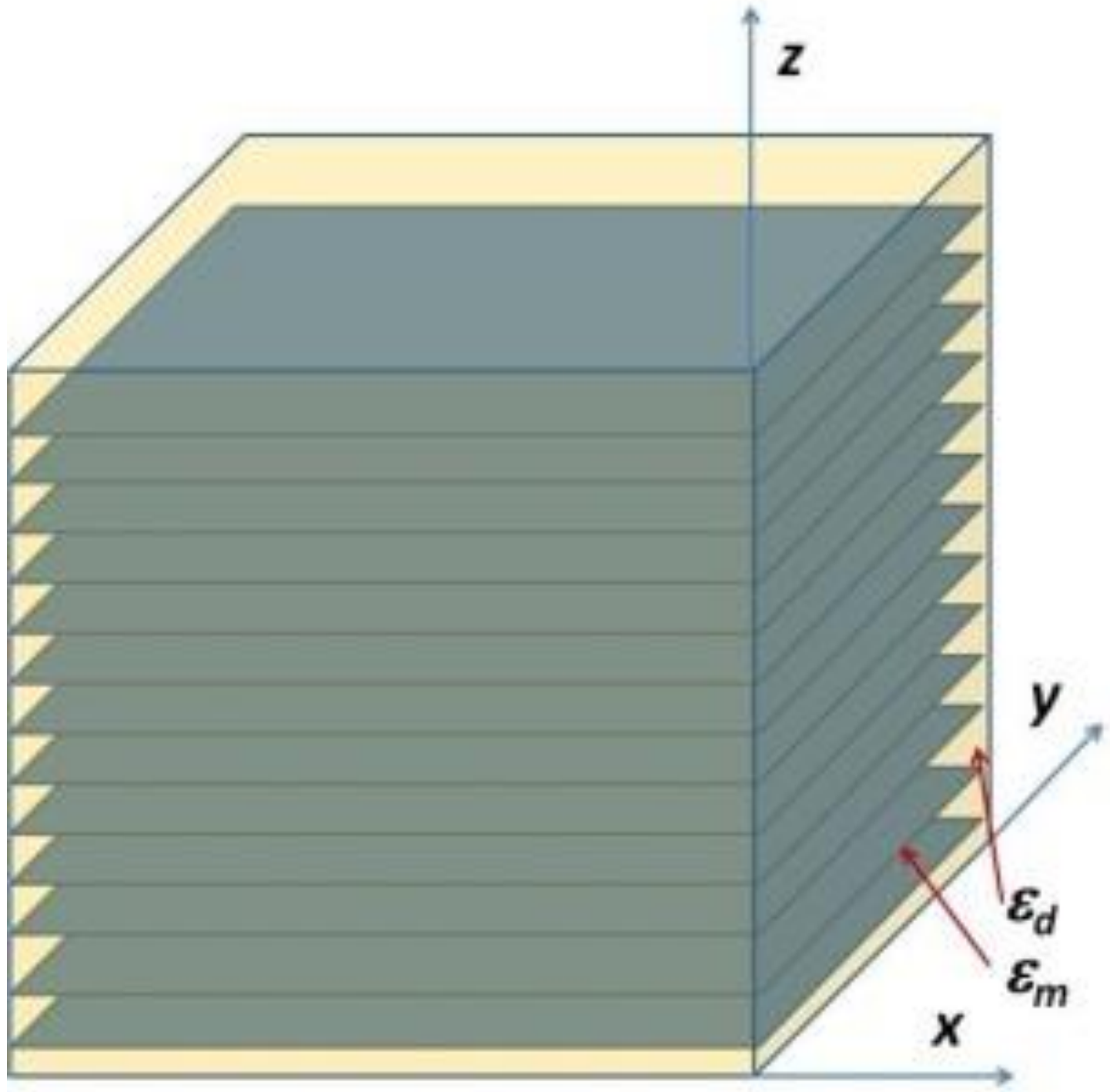

**Fig. 1**

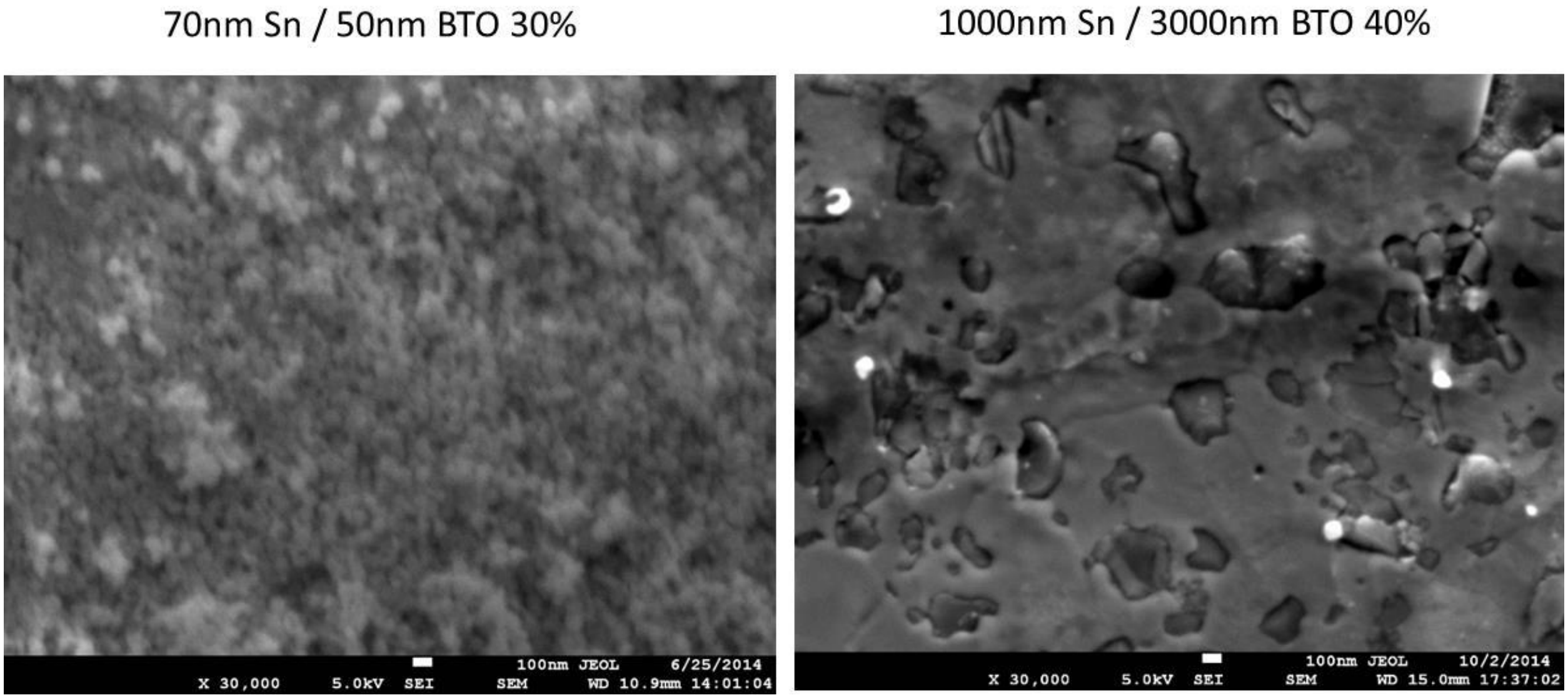
70nm Sn / 50nm BTO 30%
1000nm Sn / 3000nm BTO 40%
100nm JEOL 6/25/2014
X 30,000 5.0kV SEI SEM WD 10.9mm 14:01:04
100nm JEOL 10/2/2014
X 30,000 5.0kV SEI SEM WD 15.0mm 17:37:02

(a)

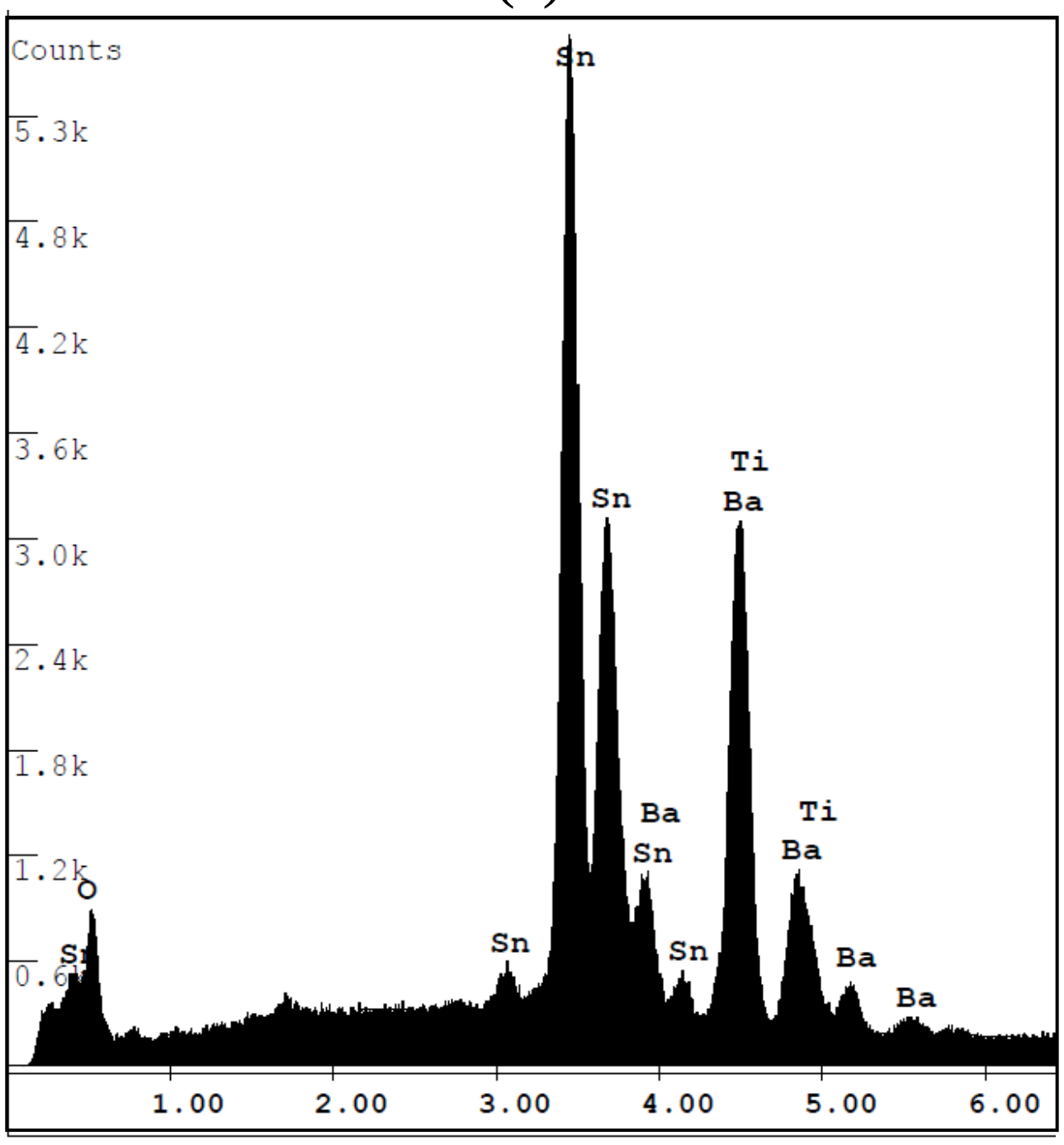

(b)

**Fig. 2**

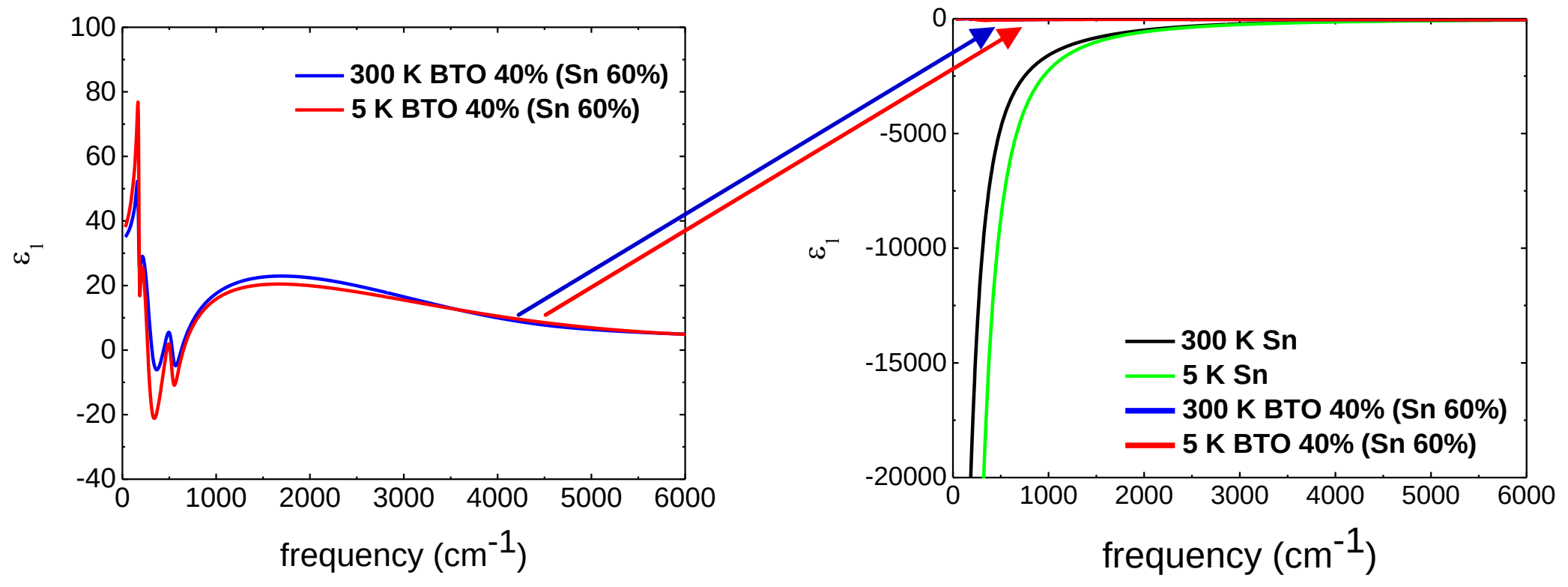

**Fig. 3**

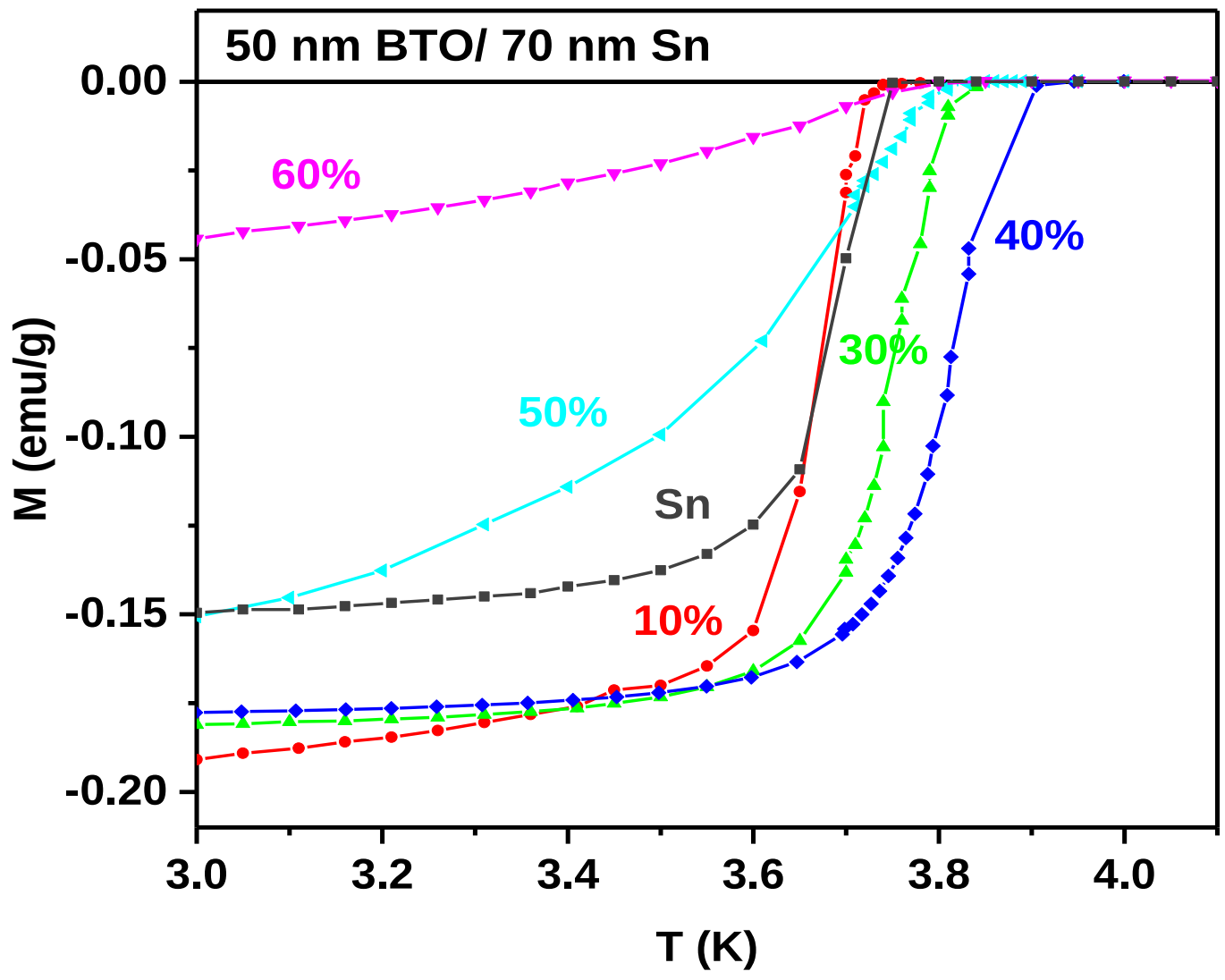

(a)

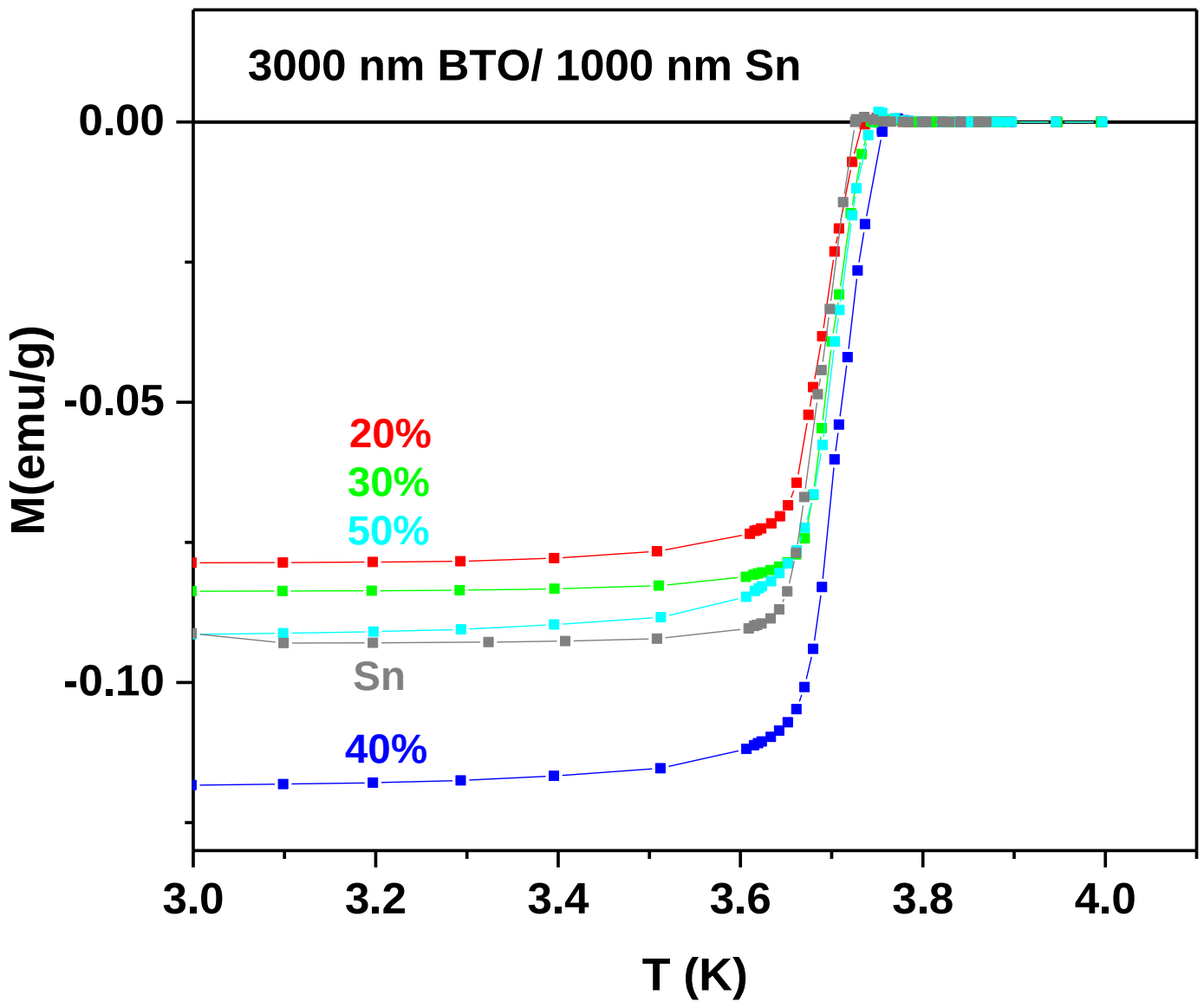

(b)

**Fig. 4**

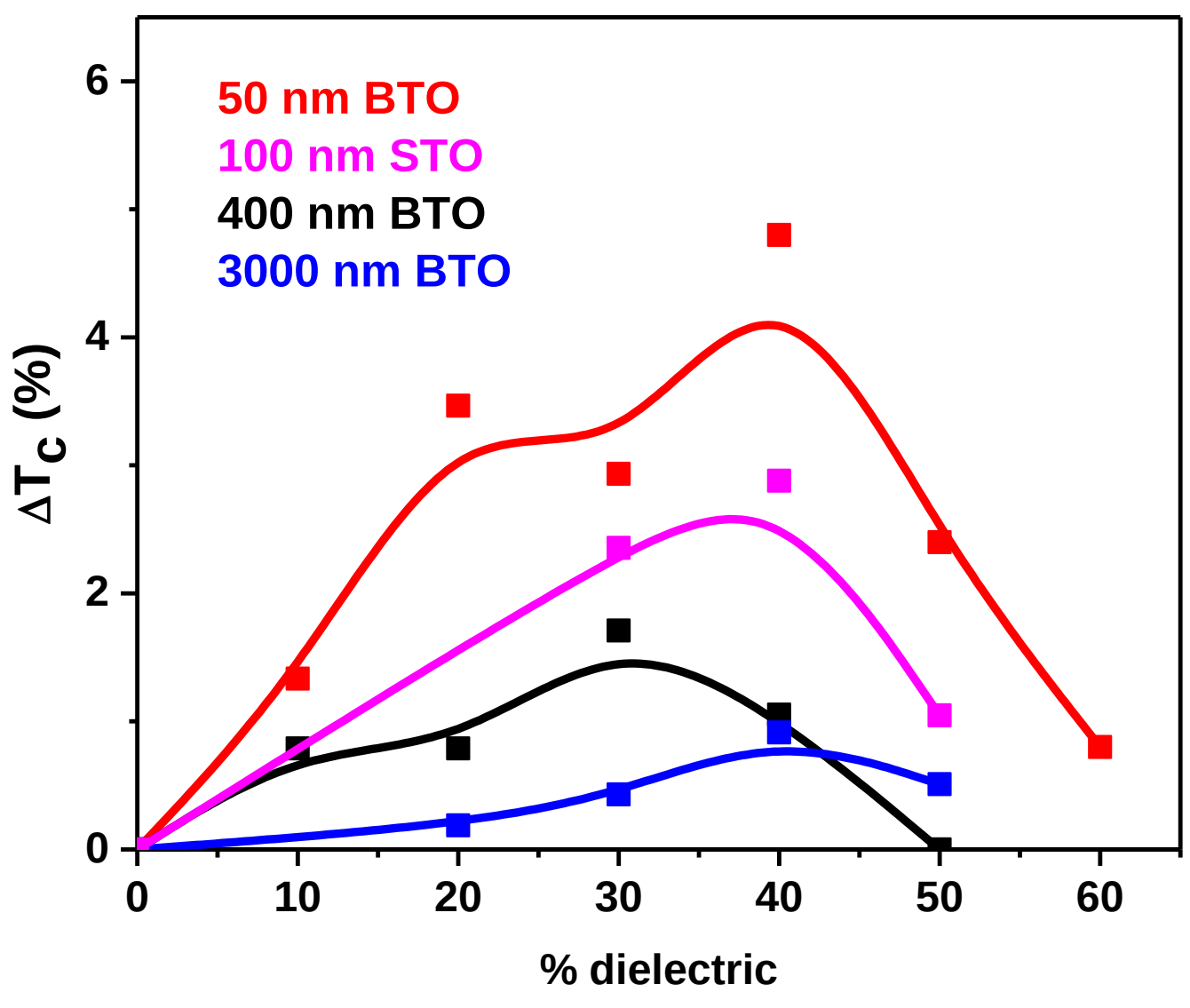

(a)

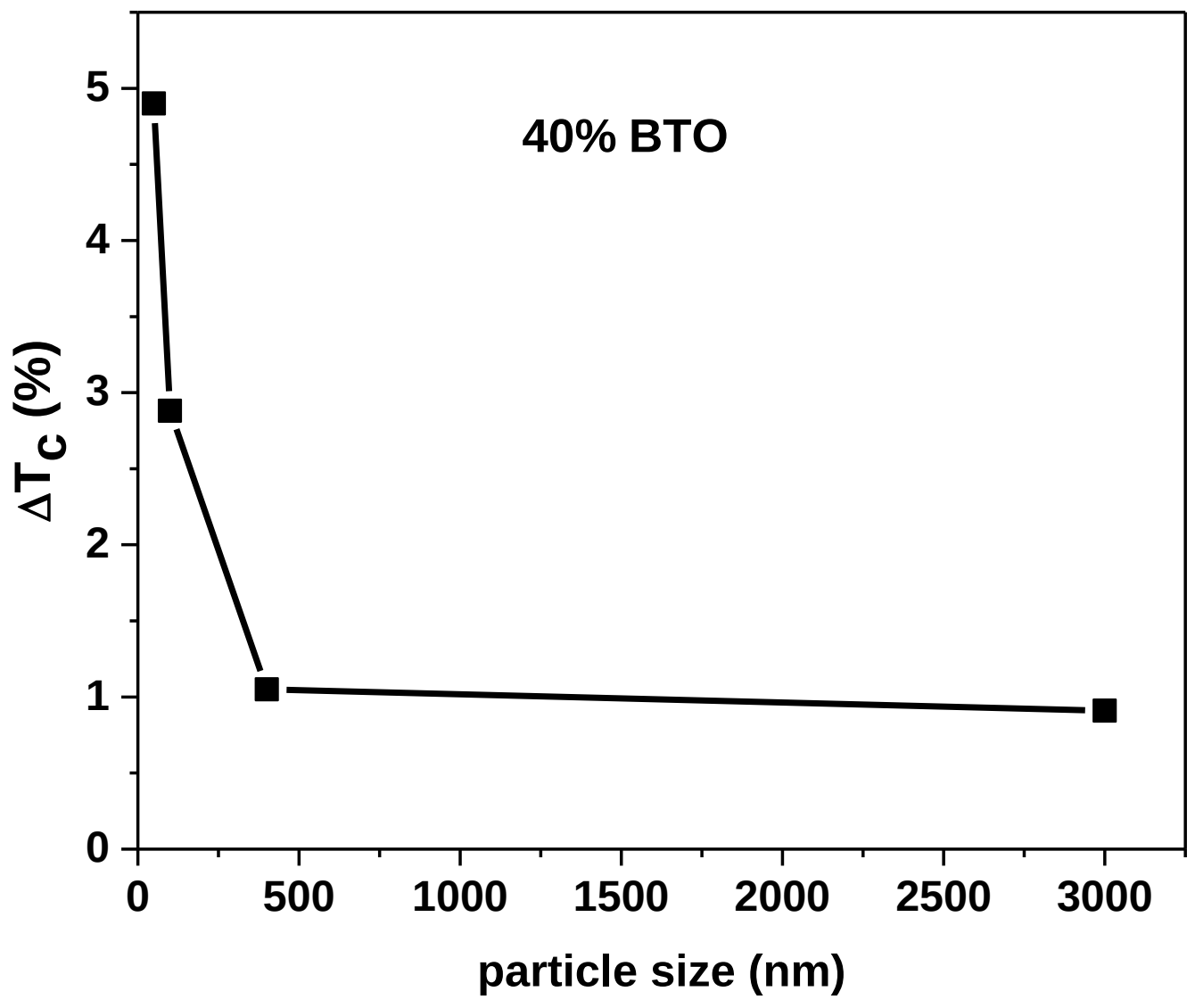

(b)

**Fig. 5**

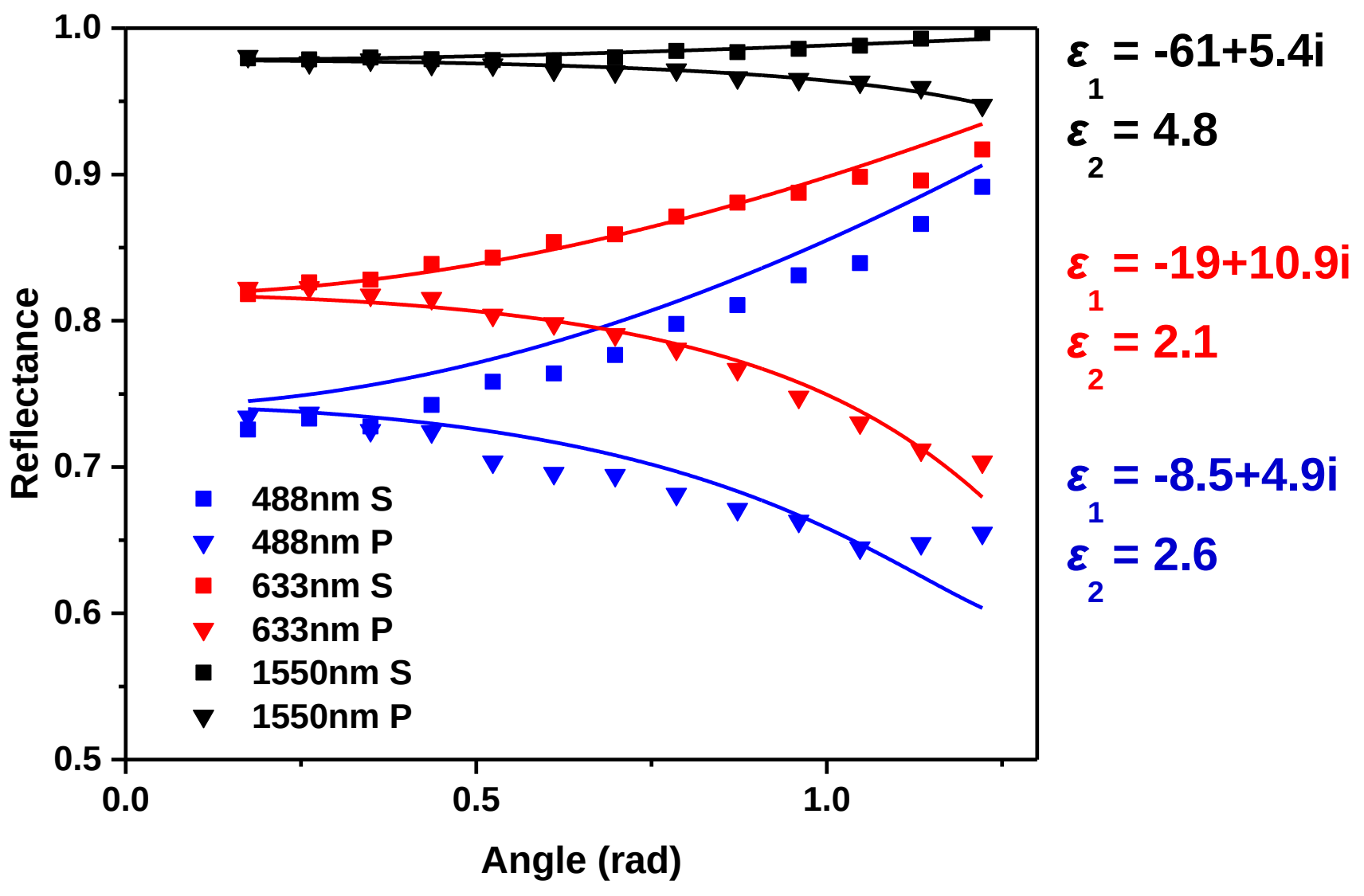

**Fig. 6**

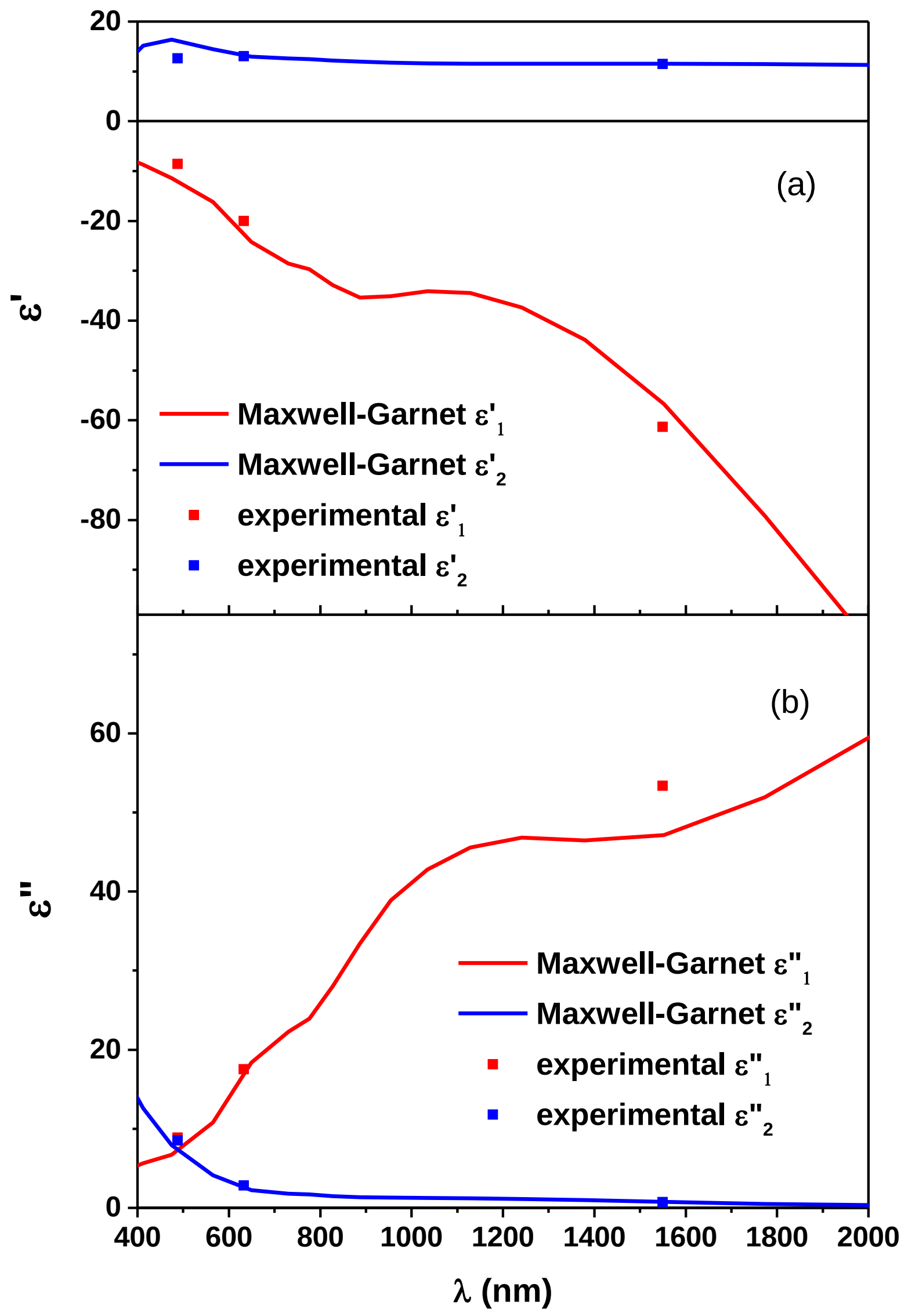

**Fig. 7**

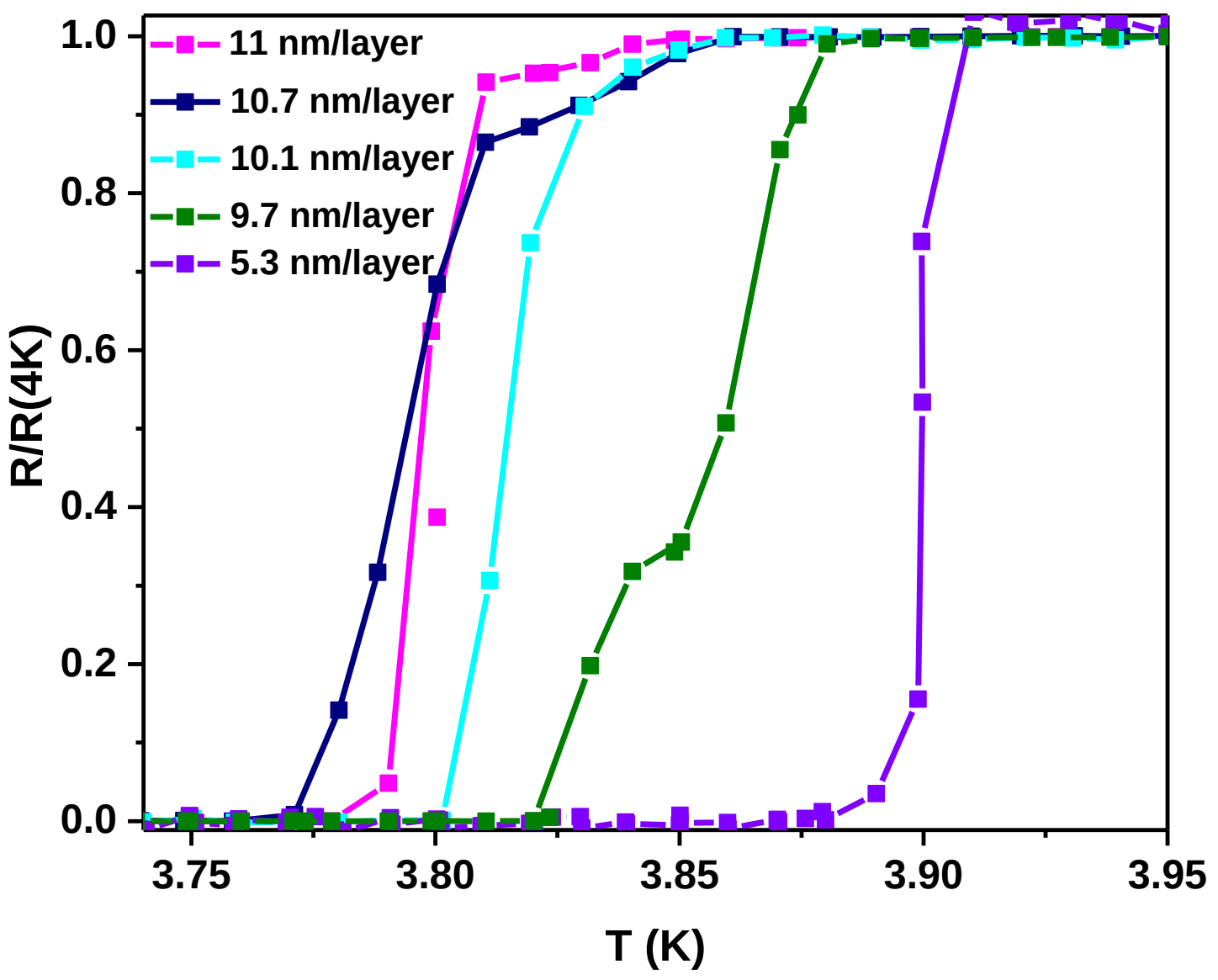

**Fig. 8**

.

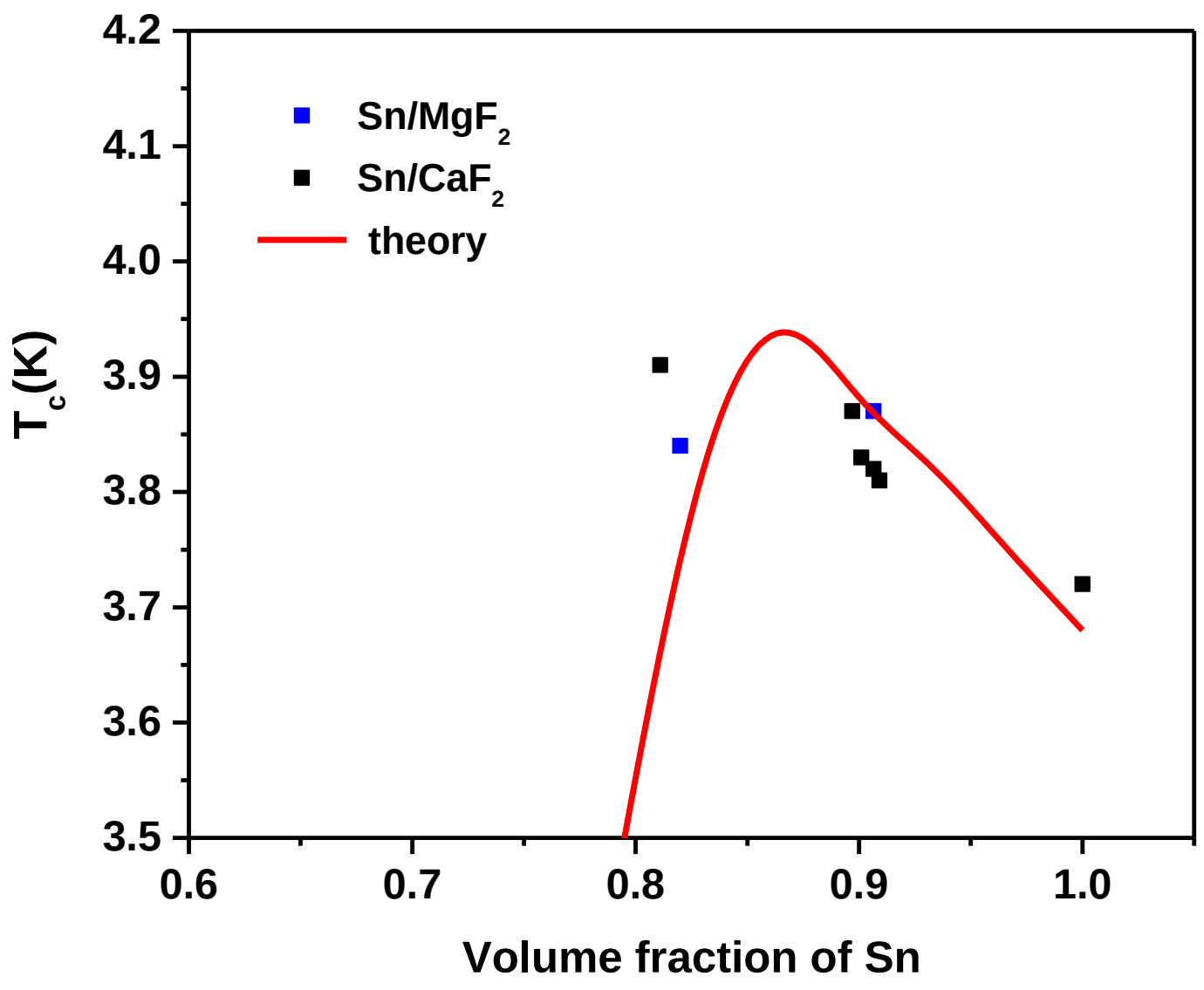

**Fig. 9**